\documentclass[12pt]{article}
\usepackage[left=2.5cm,right=2.5cm,
top=3cm,bottom=3cm]{geometry}
\usepackage[T2A]{fontenc}
\usepackage[utf8]{inputenc}
\usepackage{amsmath,amssymb,amsthm,mathtools,bm}
\usepackage{euscript,mathrsfs}
\usepackage{icomma}
\usepackage{graphicx}
\usepackage{float}
\usepackage{wrapfig}
\usepackage[dvipsnames]{xcolor}
\usepackage{indentfirst}
\usepackage{amsbsy}
\usepackage{wasysym}
\usepackage{hyperref}
\usepackage{cancel}
\usepackage{verbatim}
\usepackage{empheq}
\usepackage{adjustbox}
\usepackage{authblk}
\usepackage[page]{appendix}
\usepackage{caption}
\usepackage{subcaption}

\usepackage{tikz-cd}
\usepackage{tikz}
\usetikzlibrary{positioning}
\usetikzlibrary{calc}

\definecolor{mylightred}{RGB}{211,79,73}
\definecolor{mydarkred}{RGB}{199,44,38}
\definecolor{mylightgreen}{RGB}{78,153,67}
\definecolor{mydarkgreen}{RGB}{43,129,33}
\definecolor{mylightpurple}{RGB}{150,107,178}
\definecolor{mydarkpurple}{RGB}{126,78,160}
\definecolor{mylightblue}{RGB}{49,101,205}
\definecolor{mydarkblue}{RGB}{20,92,205}

\definecolor{urlcolor}{HTML}{120099}
\definecolor{linkcolor}{HTML}{005F5F}
\hypersetup{pdfstartview=FitH,  linkcolor=linkcolor,urlcolor=urlcolor, colorlinks=true,citecolor=blue}

\DeclareMathOperator{\arccosh}{arccosh}

\newcommand{\eps}{\epsilon}

\numberwithin{equation}{section}

\title{\textbf{Holographic two-point functions of heavy operators revisited}}
\author[]{ Prokopii Anempodistov\thanks{anempodistov.pa@gmail.com}}
\date{}

\affil[]{\small \it Laboratoire de Physique de l’École Normale Supérieure, CNRS, Université PSL, Sorbonne Universités, 
24 rue Lhomond, 75005 Paris, France}

\begin{document}

\maketitle

\begin{abstract}
    In this paper we investigate the holographic computation of the two-point functions of $\frac{1}{2}$-BPS chiral primary operators with scaling dimensions $\Delta \sim N$ or $\Delta \sim N^2$ in $\mathcal{N}=4$ $SU(N)$ SYM using Type IIB supergravity. First we consider giant graviton operators, resolving ambiguities in the previous literature on holographic computation of the two-point function, and make a new proposal for this calculation. We argue that the D3-brane action for the giant gravitons (as well as for their $\frac{1}{4}$- and $\frac{1}{8}$-BPS counterparts) should contain additional boundary terms which arise naturally from the path integral and which are required to make the variational problem well-defined. We derive the form of these terms and show that the corrected action has an on-shell value that reproduces the two-point function of the gauge theory operators. Moreover, we demonstrate that these boundary terms are necessary building blocks for the giant graviton three-point functions, and we reproduce the coordinate dependence of the extremal three-point function as a saddle-point for the boundary action. Then we consider operators with $\Delta \sim N^2$ and calculate the two-point function by evaluating the Gibbons-Hawking-York boundary term in the Type IIB pseudo-action in the Lin-Lunin-Maldacena bubbling geometry background.
\end{abstract}

\newpage

\section{Introduction}

In the AdS/CFT correspondence, which relates four-dimensional $\mathcal{N}=4$ $SU(N)$ Supersymmetric Yang-Mills theory with Type IIB string theory on asymptotically $AdS_5 \times S^5$ background \cite{Maldacena:1997re}, the holographic dictionary between the operators in the gauge theory and states in the gravity dual depends on the scaling dimensions $\Delta$ of the operators. In this dictionary the light fields with $\Delta \sim 1$ correspond in supergravity to Kaluza-Klein modes with Gubser-Klebanov-Polyakov-Witten \cite{Gubser:1998bc,Witten:1998qj} prescription for the calculation of the $n$-point functions. However, when the conformal dimension of the operator scales as $\Delta \sim N$, the KK-mode description becomes invalid, and these operators are described by extended objects in supergravity (e.g. giant gravitons, spinning string states, etc.). Moreover, since the ten-dimensional Newton constant scales as $G_{10} \sim 1/N^2$, the operators with $\Delta \sim N^2$ have sufficient energy to backreact on the geometry, and these operators should be described by new geometries. 

In this paper we will restrict our attention to the $\frac{1}{2}$-BPS sector of $\mathcal{N}=4$ SYM. Using the complex scalar $Z(x) = \Phi_1(x)+i \Phi_2(x)$, where $\Phi_{1,2}$ are two of the six scalars in SYM, one can construct a convenient basis for half-BPS chiral primaries given by Schur polynomials $\chi_R (Z)$ \cite{Corley:2001zk, Berenstein:2004kk}, where $R$ is an irreducible representation of $U(N)$ described by a Young tableau, and the number of boxes in it is equal to the conformal dimension of the operator. 
For small representation the KK-mode description is still valid, but when $\Delta \sim N$ the fully antisymmetric representation (with single column) is described by a giant graviton -- rotating spherical D3-brane that wraps $S^3$ inside $S^5$, and the symmetric representation (with single row) is described by a dual giant graviton -- rotating spherical (anti) D3-brane that wraps $S^3$ inside $AdS_5$ \cite{McGreevy:2000cw, Grisaru:2000zn, Hashimoto:2000zp, Balasubramanian:2001nh}.

It was argued in \cite{Bissi:2011dc}, following the arguments from \cite{Caldarelli:2004yk}, that the two-point function can be obtained by introducing an einbein $e$ to rewrite the action in the Polyakov formulation, then solving for the fields in terms of the $e$ and plugging it back into the action. It was argued there that after integrating the fields out the action for the einbein is the same as for the point particle, and the on-shell action then gives standard two-point function behavior. However, the usual D3-brane action, given by the sum of the Dirac-Born-Infeld and Wess-Zumino terms, is zero on-shell for the giant graviton configuration, and it should still be zero if one rewrites it using einbein, which leads one into an apparent paradox.
Another prescription for calculating the two-point functions was suggested in \cite{Bak:2011yy}, where (following the arguments from \cite{Caldarelli:2004yk}) it was claimed that one has to perform partial Legendre transformation, and the two-point function should be given by the on-shell value of the integrated Routhian. However, it is unclear why in this case one should make this partial Legendre transform, and for other types of operators \cite{Janik:2010gc, Minahan:2012fh, Klose:2011rm} this transformation is not needed. In this paper we disprove the arguments from \cite{Bissi:2011dc} and \cite{Bak:2011yy}. We show that the giant gravitons and their $\frac{1}{4}$- and $\frac{1}{8}$-BPS analogues are required to have additional boundary terms in their actions, derive the form of these boundary terms and as a consistency check we show that their on-shell value reproduces the two-point function behavior. Moreover, we show that the boundary term action gives a well-defined variational problem to tackle higher-point functions. For example, we reproduce the extremal three-point function coordinate behavior by considering a "three-legged" configuration consisting of three D3-branes (which wrap the same $S^3 \subset S^5$)
meeting at a common point in the bulk by minimizing the boundary term action over this meeting point. 

As was mentioned earlier, operators with $\Delta \sim N^2$ sufficiently backreact on the background and should be described by deformed geometries. For the generic type of operators (non-supersymmetric) the two-point function was recovered by evaluating the on-shell action for the AdS-Schwarzschild geometry in \cite{Abajian:2023jye,Abajian:2023bqv}. For the two-point functions of half-BPS operators one expects the supergravity background to be also half-BPS, and the most general non-singular background which preserves half of the supersymmetries and has $R \times SO(4) \times SO(4) $ symmetry of the two-point function were
constructed by Lin, Lunin, and Maldacena in \cite{Lin:2004nb} (for brevity, we call them LLM backgrounds in the following). 
One can verify that the LLM background is, in fact, the geometry created by the insertion of two heavy ($\Delta \sim N^2$) half-BPS operators by performing a probe analysis: one inserts a third light ($\Delta \sim 1$) operator on the boundary and does GKPW analysis in the LLM background to extract the three-point function and compares with the CFT computation, as was first done in \cite{Skenderis:2006uy}.
Another way of obtaining LLM geometry through identifying its boundary droplet as an eigenvalue distribution of a complex matrix model, as well as calculating various three-point functions, have been proposed in \cite{Anempodistov:2025maj} (see also \cite{Kazakov:2024ald}). However, to the best of our knowledge there were no calculations of the two-point function of heavy half-BPS operators directly from the ten-dimensional supergravity in the LLM background. This calculation might also be relevant for the computation of the three-point functions \cite{Holguin:2025dei} and four-point functions and integrated correlators \cite{Aprile:2024lwy,Aprile:2025hlt,Turton:2024afd,Turton:2025svk, Turton:2025cnn, Aprile:2026urq}.

In this paper we address the problem of calculating the two-point functions in these two regimes ($\Delta \sim N$ and $\Delta \sim N^2$) directly from the supergravity. One could rightfully argue that the two-point functions may be calculated just by evaluating the energy of the configuration, but of course, the goal of this is not just simply reproducing the right behavior of the two-point functions. We believe that consistent calculation of the two-point functions in the worldline formalism is a necessary prerequisite to tackle the problem of gravity calculation of three-point functions (where the energy argument is not applicable). For example, we derive here the boundary terms for the giant graviton configurations (whose presence is not a choice, but a necessity because they ensure that the variation of the action over the fields goes to zero on-shell), and we show that only with these boundary terms the action is non-zero on-shell.  Then, the presence of these boundary terms allows one to set up a nice variational problem for the three-point function calculation. For example, for the extremal correlator one may consider a configuration that consists of three geodesics meeting in the bulk, and then one can minimize the action of this configuration over the joining point to obtain the three-point function \cite{Minahan:2012fh,Klose:2011rm}. We carry out this calculation and show that the boundary term action gives the correct coordinate dependence of the extremal three-point function.  
We expect that, in a semiclassical saddle-point description, these boundary terms control the spacetime dependence of non-extremal three-point functions, while the full structure constant requires the dynamics of the brane in the internal directions.

The outline of this paper is as follows: In Section \ref{sec-GG} we show that the D3-brane action for the giant gravitons and dual giant gravitons should contain additional boundary terms that make the variational problem well-defined and which arise naturally from the path-integral. We show that the on-shell value of the corrected action reproduces the standard behavior of the two-point function, and the whole contribution comes from these boundary terms. We also comment on why these boundary terms are necessary for the three-point functions and calculate the coordinate dependence of the extremal three-point function. Then, in Section \ref{sec-LLM} we calculate the two-point function for operators with $\Delta \sim N^2$ by evaluating the Type IIB pseudo-action in the LLM background. We show that in complete analogy with the giant graviton case, the bulk action vanishes on-shell and the two-point function behavior comes from the Gibbons-Hawking-York boundary term. Appendix \ref{app-GG} contains review of the arguments in the literature about the giant graviton two-point function, and we explain where these arguments break down. Appendix \ref{app-gg-eom-poincare} contains some technical details about motion of giant gravitons in Poincar\'e coordinates.  Appendix \ref{app_asymptotic-expansions} contains the details of the asymptotic expansion of the LLM metric near the boundary. 
Finally, in Appendix \ref{app-thin-ring} we demonstrate that the (dual) giant graviton two-point functions can equally be calculated using LLM geometries with thin ring distributions.

\section{Giant graviton and dual giant graviton two-point functions} \label{sec-GG}

\subsection{Giant gravitons in $S^3 \subset S^5$ and derivation of the boundary terms}
Let us start the discussion with the giant graviton case.
The gauge theory correlator we are interested in is \cite{Corley:2001zk}:
\begin{equation} \label{GG-2pt}
    \langle \chi_{(1^k)}(Z(x_1)) \chi_{(1^k)}(\bar{Z}(x_2)) = \frac{1}{|x_1-x_2|^{2k}} \times \prod_{i=1}^k (N-i+1),
\end{equation}
with $\chi_R(Z)$ being a Schur polynomial, and where $(1^k)$ is an antisymmetric representation corresponding to a single column Young tableau with $k$ boxes. The normalization of the states can be obtained using prescription from \cite{Holguin:2025dei} and the goal of this paper is to reproduce the coordinate behavior of this two-point function, which we will do by evaluating the on-shell action for giant gravitons. 

We choose the following coordinates for $AdS_5 \times S^5$ metric:
\begin{equation}
\begin{aligned}
       ds^2& = -\cosh^2 \rho dt^2 + d\rho^2 + \sinh^2\rho \left(d\alpha_1^2 + \sin^2 \alpha_1 \left(d\alpha_2^2 + \sin^2\alpha_2 d\alpha_3^2\right)\right)\\
       +& d\theta^2 +\cos^2 \theta d \varphi^2 + \sin^2\theta\left(d\beta_1^2 + \sin^2 \beta_1 \left(d\beta_2^2 + \sin^2\beta_2 d\beta_3^2\right)\right).
\end{aligned} \label{AdS-metric}
\end{equation}

The gravity dual of \eqref{GG-2pt} is D3-brane with the following embedding \cite{Grisaru:2000zn,Hashimoto:2000zp,Balasubramanian:2001nh, Bissi:2011dc}:
\begin{equation} \label{GG-embedding}
    \sigma^0 =t, \quad \sigma^i = \beta_i, \quad \varphi = \varphi(t), \quad \rho=0.
\end{equation}
The action is given by a sum of the Dirac-Born-Infeld and Wess-Zumino terms, and after integrating out the $S^3$ variables, one obtains \cite{Grisaru:2000zn,Bissi:2011dc}:
\begin{multline} \label{GG-action}
    S_{D3} = S_{DBI}+S_{WZ}= -\frac{N}{2\pi^2} \bigg( \int d^4 \sigma \sqrt{-g} - \int P[C_4] \bigg)=  \\
    =\int dt \, L=-N \sin^3 \theta \int dt \sqrt{1- \cos^2 \theta \dot{\varphi}^2}+N\sin^4 \theta \int \dot{\varphi} dt ,
\end{multline}
where $P[C_4]$ is the pullback of the 4-form potential $C_4$ on the brane worldvolume. This action possesses a conserved angular momentum
\begin{equation}
    k \equiv \frac{\partial L}{\partial \dot{\varphi}} = \frac{N \sin^3 \theta \cos^2 \theta  \dot{\varphi}}{\sqrt{1-\cos^2 \theta \dot{\varphi}^2}} +N \sin^4 \theta .
\end{equation}
The solution of equations of motion is
\begin{equation}
    \dot{\varphi}=1,
\end{equation}
and the expression for the angular momentum simplifies
\begin{equation}
    k = N \sin^2 \theta,
\end{equation}
and by plugging this solution into the action \eqref{GG-action} one can easily see that it evaluates to zero:
\begin{equation} \label{DBI-plus-WZ-zero}
    (S_{DBI}+S_{WZ})_{|\text{on-shell}}=0.
\end{equation}
The fact that the sum of DBI and WZ terms goes to zero is not a coincidence, but is tightly related to the fact that the brane embedding satisfies the kappa-symmetry, and hence, is supersymmetric \cite{Barwald:1999hx,Dadok,Simon:2011rw,Harvey,Koerber:2005qi,Martucci:2005ht,Papadopoulos:2006hk}. The reason is that the kappa-symmetry projection condition
\begin{equation}
    \Gamma \eps = \eps,
\end{equation}
where $\eps$ is a Killing spinor of $AdS_5 \times S^5$ and $\Gamma$ is a kappa-symmetry projector, which for the D3-branes without gauge field excitations can be written as (see, e.g. \cite{Skenderis:2002vf} for the general form):
\begin{equation}
    \Gamma = \frac{1}{\sqrt{-g}} \frac{1}{4!} \varepsilon^{i_1 ...i_4} \partial_{i_1}X^{m_1} ... \partial_{i_4} X^{m_4}\, \gamma_{m_1...m_4},
\end{equation}
where $\{i_k\}$ indices run in the brane worldvolume coordinates, $\{m_k\}$ run in the target space coordinates, and $\gamma_i$ are curved spacetime gamma matrices. Then, roughly by multiplying this condition on the left by $\eps^{\dagger}$ one can show that it is equivalent to saturation of the BPS bound, which may be written as
\begin{equation}
     \sqrt{-g} = \frac{1}{4!} \varepsilon^{i_1...i_4}\partial_{i_1}X^{m_1} ... \partial_{i_4} X^{m_4}\, C_{m_1...m_4},
\end{equation}
which leads to \eqref{DBI-plus-WZ-zero} for any supersymmetric embedding of the brane\footnote{It can be straightforwardly verified that the embedding \eqref{GG-embedding} with $\dot{\varphi}=1$ preserves half of the supersymmetry \cite{Grisaru:2000zn,Hashimoto:2000zp}.}. In principle, this fact would lead to a problem for the gravity calculation of the three-point function of the giant gravitons. Since this three-point function in the gauge theory is $\frac{1}{4}$-BPS, it is reasonable to expect that in the gravity dual it would correspond to a $\frac{1}{4}$-BPS configuration of the D3-brane with a "pair-of-pants" structure, and the exact profile of this brane embedding should be found by minimizing its on-shell action (as it is done for the short string states \cite{Minahan:2012fh} or BMN strings \cite{Klose:2011rm}). However, since for any $\frac{1}{4}$-BPS configuration one would have that the sum of DBI and WZ terms cancel, there would be nothing to minimize and one would run into a pathology. 
However, below we show that when one does holographic computation, the action for a giant graviton should also contain additional boundary terms, and the fact that DBI and WZ terms compensate each other does not lead to a conclusion that the on-shell action is zero.

Now we make a new proposal on calculation of the two-point functions of giant gravitons.
The idea is that the two-point function is still given by the on-shell value of the action, but the action itself should be modified by the addition of the boundary terms that arise from the boundary conditions: in the functional integral for the two-point function
\begin{equation} \label{path-int-2pt}
    \mathcal{Z}_{2pt} = \int \mathcal{D}\varphi \exp \bigg(-N \sin^3 \theta \int dt \sqrt{1- \cos^2 \theta \dot{\varphi}^2}+N\sin^4 \theta \int \dot{\varphi} dt \bigg),
\end{equation}
we should integrate over $\varphi$ with Neumann boundary conditions. Said differently, when we compute the two-point function, we implicitly impose the condition that at the asymptotic infinities the angular momentum is equal to $k$, and we should add this condition directly into the action. This will also guarantee that the variation of the action will vanish: 
\begin{equation}
    \delta S = \int^{t_f}_{t_i} dt \frac{\partial L}{\partial \dot{\varphi}} \delta \dot{\varphi} = k \, \delta \varphi \Big|^{t_f}_{t_{i}} \neq 0,
\end{equation}
where we have used that
\begin{equation}
    k = \frac{\partial L}{\partial \dot{\varphi}} =    \frac{N \sin^3 \theta \cos^2 \theta \dot{\varphi}}{\sqrt{1-\cos^2 \theta \dot{\varphi}^2}} +N \sin^4 \theta  = \text{constant on-shell}.
\end{equation}
Since we integrate over $\varphi$ with the Neumann boundary conditions this variation does not vanish unless we add a boundary term\footnote{Similar terms were also discussed in \cite{Drukker:2005kx} for the D3-branes describing Wilson loops. For the case of the giant gravitons, these terms have a different role, which we discuss below.} to the action as follows:
\begin{equation} \label{boundary-terms}
    S_N = S - \, k \,{\varphi(t)}\Big|^{t_f}_{t_{i}}.
\end{equation}
When we calculate the on-shell value of the action, the "bulk" part will be equal to zero as shown above, and the whole contribution will come from the boundary terms:
\begin{equation} \label{action-gg-onshell}
    S_N^{\text{on-shell}} = -k (t_f-t_i) = i k (\tau_f-\tau_i),
\end{equation}
where we have done Wick rotation $t = -i \tau$. 
Now, let's determine $\tau_f$ and $\tau_i$. The $\rho =0$ line in the global $AdS$ maps to a semicircle in the Poincaré coordinates:
\begin{equation} \label{Poincare-geodesic}
    \begin{aligned}
        z = \frac{R}{\cosh \tau },\qquad 
        x_0^E= R\tanh \tau ,
    \end{aligned}
\end{equation}
so the global time becomes a parameter along the geodesic. Then, because there is a cutoff at $z=\eps$, in the global $AdS$ picture there is also a cutoff at large $t$ (or $\tau$), as shown in Fig. \ref{fig:GG}:
\begin{figure}[H]
    \centering
    \includegraphics[width=0.8\linewidth]{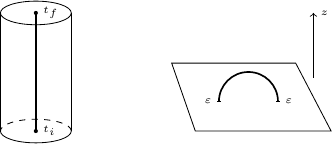}
    \caption{The giant graviton profile in the global $AdS$ (left) and the same profile in the Poincaré coordinates (right). We emphasize here that the $\rho=0$ worldline in the global coordinates translates into the semicircular geodesic when translated into Poincaré, and the cutoffs at $z=\eps$ are naturally translated into cutoffs for the global time.}
    \label{fig:GG}
\end{figure}
Plugging $z=\eps$ into \eqref{Poincare-geodesic}, we find
\begin{equation}
    \tau_f = \text{arccosh} \frac{R}{\eps}, \qquad \tau_i = -\text{arccosh} \frac{R}{\eps},
 \end{equation}
and using the fact that $R = |x_1-x_2|$ we obtain for the on-shell action
\begin{equation}
    i S_N^{\text{on-shell}} = -2k \arccosh \frac{R}{\eps} \approx -2k \log \frac{2R}{\eps} = -2k \log \frac{|x_1-x_2|}{\eps},
\end{equation}
and we obtain the standard coordinate behavior of the two-point function \eqref{GG-2pt}:
\begin{equation} \label{gg-2pt}
    \langle\mathcal{O}_{k}(x_1) \mathcal{O}_{k} (x_2) \rangle_{} \sim \bigg( \frac{\eps}{|x_1-x_2|} \bigg)^{2k},
\end{equation}
where $\mathcal{O}_k(x) \equiv \chi_{(1^k)}(Z(x))$.

\subsection{Generalization to dual giant gravitons in $S^3 \subset AdS_5$ and $\frac{1}{4}$- and $\frac{1}{8}$-BPS cases}

Now we generalize the derivation of the boundary terms discussed above for the case of the dual giant graviton which wraps $S^3$ inside $AdS_5$. 

The gauge theory correlator in this case is
\begin{equation} \label{dualGG-2pt}
    \langle \chi_{(k)}(Z(x_1)) \chi_{(k)}(\bar{Z}(x_2)) = \frac{1}{|x_1-x_2|^{2k}} \times \prod_{i=1}^k (N-1+i),
\end{equation}
where $(k)$ is a symmetric representation with Young tableau consisting of a single row of length $k$. This is dual to a (anti)D3-brane with the following action \cite{Grisaru:2000zn, Hashimoto:2000zp,Bissi:2011dc}:
\begin{equation}
    S_{D3} = -N \int dt \big( \sinh^3\rho \sqrt{\cosh^2 \rho - \dot{\varphi}^2} - \sinh^4 \rho  \big),
\end{equation}
which is again solved by $\dot{\varphi}=1$, and the angular momentum is
\begin{equation}
    k = \frac{N \dot{\varphi} \sinh^3 \rho}{\sqrt{\cosh^2 \rho - \dot{\varphi}^2}} = N \sinh^2 \rho. 
\end{equation}
Since we are again fixing the angular momentum on the endpoints, we add the same boundary term into the action $S_N = S - \, k \,{\varphi(t)}\Big|^{t_f}_{t_{i}}$, and the on-shell action is again
\begin{equation}
    S_N^{\text{on-shell}} =  i k (\tau_f-\tau_i).
\end{equation}
We again need to find how $\tau_f$ and $\tau_i$ are related to the insertion points on the boundary.
Now the brane sits at the $\rho=\text{const}$ cylinder in the global $AdS$, and the transformation from the global coordinates to Poincaré is:
\begin{equation} \label{global-to-poincare}
    \begin{aligned}
     z = \frac{R}{\cosh \rho \cosh \tau - n_0 \sinh \rho},\\
     x^0_E = \frac{R \cosh \rho \sinh \tau}{\cosh \rho \cosh \tau - n_0 \sinh \rho},\\
     \Vec{x} = \frac{R \, \Vec{n} \sinh \rho}{\cosh \rho \cosh \tau - n_0 \sinh \rho},
     \end{aligned}
\end{equation}
where $(n_0,\Vec{n})$ is the unit normalized vector on $S^3$ in $4$-dimensional embedding space. The structure of the $\rho=\text{const}$ cylinder in the Poincaré coordinates is shown in Fig. \ref{fig:dualGG}:
\begin{figure}[H]
    \centering
    \includegraphics[width=0.6\linewidth]{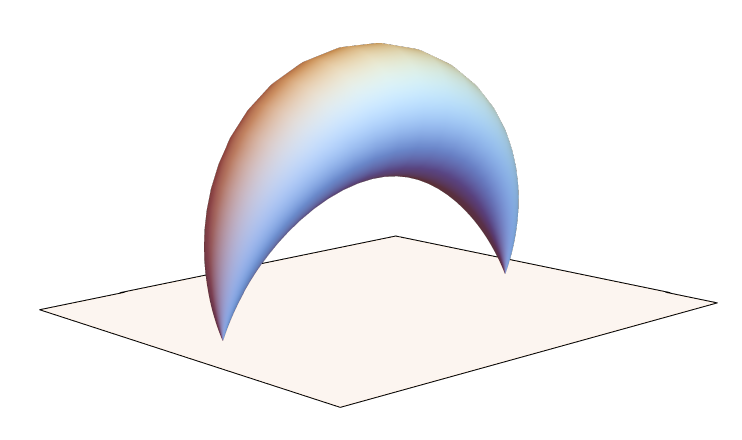}
    \caption{The dual giant graviton profile in the Poincaré coordinates.}
    \label{fig:dualGG}
\end{figure}
The boundary points correspond to $\tau \gg 1$, and in this limit we can neglect the second terms in the denominators in \eqref{global-to-poincare}. Then, we find
\begin{equation}
    \tau_f = \text{arccosh} \bigg(\frac{R}{\eps \cosh \rho}\bigg), \qquad \tau_i = -\text{arccosh} \bigg(\frac{R}{\eps \cosh \rho}\bigg).
\end{equation}
The additional factor of $\cosh \rho$ adds a coordinate-independent term in the on-shell action,
which can be neglected\footnote{See \cite{Holguin:2025dei} for the discussion of the normalization of the two-point functions.}, and the two-point function reproduces \eqref{dualGG-2pt}:
\begin{equation}
    \langle \mathcal{O}_{k}(x_1) \mathcal{O}_{k} (x_2) \rangle \sim \bigg( \frac{\eps}{|x_1-x_2|} \bigg)^{2k},
\end{equation}
where $\mathcal{O}_k(x) \equiv \chi_{(k)}(Z(x))$.

The arguments above can be straightforwardly generalized to the case of $\frac{1}{4}$- and  $\frac{1}{8}$-BPS giant gravitons. Such solutions can be generated using Mikhailov's construction \cite{Mikhailov:2000ya} as follows: if one considers $S^5$ to be embedded in $\mathbb{C}^3$ with flat complex coordinates $Z_i$ (where $i=1,2,3$ and $Z_i$ are identified with the complex scalars from SYM), then the $t=0$ slice of the worldvolume of the giant graviton can be found as an intersection of the $S^5$ with a given holomorphic surface:
\begin{equation}
    \begin{cases}
            |Z_1|^2+|Z_2|^2+|Z_3|^2=1,\\
            F(Z_1,Z_2,Z_3)=0,
    \end{cases}
\end{equation}
and the full worldvolume can be found by translating
\begin{equation} \label{Mikhailov-time-evolution}
    \{Z_1,Z_2,Z_3\}\to \{e^{it}Z_1,\,e^{it}Z_2,\,e^{it}Z_3 \}.
\end{equation}
It can be shown that for a general holomorphic surface $F(Z_1,Z_2,Z_3)$ the brane embedding preserves $\frac{1}{8}$ of the supersymmetries, but for holomorphic surfaces with fewer scalars $F(Z_1,Z_2)$ and $F(Z_1)$ the preserved supersymmetries enhance to $\frac{1}{4}$ and $\frac{1}{2}$ respectively \cite{Mikhailov:2000ya}. 

The boundary term in this case can be derived straightforwardly in complete analogy with the $\frac{1}{2}$-BPS case described above. Below we repeat some arguments from \cite{Mikhailov:2000ya,Hackett-Jones:2004hay} to demonstrate how the boundary terms arise in this case and how their on-shell value reproduces the coordinate dependence of the two-point function. 

One starts by defining a one-form $e^{\perp} \in T^* \mathbb{C}^3$ which is orthogonal to the $S^5$, and defining another one-form $e^{\parallel} \in T^*S^5$ by the action with the complex structure: $e^{\parallel} = I. e^{\perp}$. 
Since the spatial worldvolume $\Sigma$ of the giant graviton is not orthogonal to $e^{\parallel}$, one can decompose the one-form $e^{\parallel}$ into normal and parallel to the $\Sigma$ components $e^{\phi}$ and $e^{\psi}$ respectively:
\begin{equation}
    e^{\parallel} = - v \,e^{\phi}- \sqrt{1-v^2} \,e^{\psi}.
\end{equation}
Now to calculate the giant graviton action, it is convenient to use the following coordinates on $S^5$:
\begin{equation} \label{metric-s5-mikhailov}
    ds^2_{S^5} = (e^{\phi})^2 + (e^n)^2+d\Sigma^2,
\end{equation}
where $e^n$ is a unit one-form on $S^5$ orthogonal to both $e^{\phi}$ and $\Sigma$. 
Then, the action of the D3-brane is \cite{Hackett-Jones:2004hay}:
\begin{equation}
    S = \frac{N}{2\pi^2}\int dt \, d\Omega_3 \big( - \sqrt{(1-\dot{\phi}^2)g_{\Omega_3}} + C_{t \sigma^1 \sigma^2 \sigma^3 } + \dot{\phi} \,C_{\phi \sigma^1 \sigma^2 \sigma^3}  \big),
\end{equation}
where $g_{\Omega_3}$ is the determinant of the metric on $\Sigma$.  
Due to the form of the metric \eqref{metric-s5-mikhailov}, one can choose a gauge for the components of $C^{(4)}$ such that they are independent of $\phi$, and one can see that the action again depends only on $\dot{\phi}$ with no explicit $\phi$ dependence. From the Mikhailov's construction $\dot{\phi}=v$, and one can show that only for this $\dot{\phi}$ the brane wraps a calibrated cycle \cite{Hackett-Jones:2004hay}. Then, one can introduce the boundary terms that fix the Neumann boundary condition in complete analogy with the $\frac12$-BPS case considered above, i.e. one adds $-k \,\phi(t) |^{t_f}_{t_i}$ to the action, where $k \equiv \frac{\partial L}{\partial \dot{\phi}}$ is constant on-shell. Evaluating this boundary term on-shell (which is the same as \eqref{action-gg-onshell}), one obtains correct coordinate dependence of the two-point function \eqref{gg-2pt}. Note, that in Mikhailov's construction the giant gravitons can have up to three angular momenta $(J_1,J_2,J_3)$ depending on the amount of preserved supersymmetries and which are associated to the $SO(6)$ rotations of $Z_i$. The fact that the brane moves with the speed of light along $e^{||}$ together with the fact that the time evolution is given by \eqref{Mikhailov-time-evolution} leads to the conclusion that $k=J_1+J_2+J_3$ for generic $\frac{1}{8}$-BPS giant graviton, where $k$ is defined above. 
Equivalently, for a configuration with several independent charges one could write the boundary terms as $-\sum_i J_i \int dt \, \dot{\phi_i}$, where $\phi_i$ are defined as $Z_i = r_i e^{i \phi_i}$, and on the BPS trajectory generated by common rotation $Z_i \to e^{it} Z_i$ this reduces to $-(\sum_i J_i)(t_f-t_i)$.
We stress that the on-shell value of the boundary term gives only a coordinate dependence of the two-point function and does not determine the normalization, which is a non-trivial function of charges $J_i$, but also depends on the precise definition of the operator in the gauge theory.

\subsection{The boundary terms as a setup for the giant graviton three-point functions}

In this section we explain why the boundary terms for the giant gravitons introduced above are essential for the three-point function calculation. We will show that variation of the on-shell value of these boundary terms over the meeting point is well-defined and gives the right saddle which reproduces the coordinate behavior of the three-point function. 

First, as a basic check let us consider two D3-branes emanating from the insertion points on the boundary and meeting at some arbitrary point in the bulk $(\Vec{x}_c \, ,\,z_c)$ in the Poincar\'e coordinates, then minimize the action coming from the boundary terms and check that the saddle point solution recovers the usual semicircular geodesic corresponding to the two-point function. 

Since both D3-branes wrap the same $S^3 \subset S^5$, their worldvolumes in the AdS part are just given by the geodesics (equations of motion coming from the D3-brane action reduce to geodesic equations when one enforces $\dot{\varphi}=1$, see Appendix \ref{app-gg-eom-poincare} for the details).
For the first geodesic let us denote the solution as
\begin{equation} \label{1st-geodesic}
    \begin{aligned}
        &x(\tau) = x_0^{(1)} + R_1 \tanh(\tau-\tau_0),\qquad 
        &z(\tau) = \frac{R_1}{\cosh (\tau-\tau_0)}.
    \end{aligned}
\end{equation}
We parametrize the geodesics by the global time $t$ and perform the Wick rotation $t=-i\tau$. We make the first geodesic begin at the boundary point $(x_1\,,\,\eps)$ at euclidean time $\tau_{in}$ and end at $({x}_c \, ,\,z_c)$ point in the bulk at time $\tau_c$.
From the boundary condition at $z=\eps$ we obtain
\begin{equation}
    \begin{aligned}
        \tau_{in} = \tau_0 - \arccosh \frac{R_1}{\eps},\qquad 
        x_0^{(1)} = x_1 + R_1,
    \end{aligned}
\end{equation}
and the conditions at $z=z_c$ give
\begin{equation}
    \begin{aligned}
        &R_1 = z_c \cosh (\tau_c-\tau_0),\qquad 
        & \tau_c=\tau_0+\log \bigg( \frac{x_c-x_1}{z_c} \bigg).
    \end{aligned}
\end{equation}
We parametrize the second geodesic as
\begin{equation}
    \begin{aligned}
        &x(\tau) = x_0^{(2)} + R_2 \tanh(\tau-\Tilde{\tau}_0),\qquad 
        &z(\tau) = \frac{R_2}{\cosh (\tau-\Tilde{\tau}_0)},
    \end{aligned}
\end{equation}
and make it start from the $({x}_c \, ,\,z_c)$ point in the bulk at time $\tau_c$ and end on the boundary $(x_2\,,\, \eps)$ at time $\tau_{out}$.
The boundary conditions give
\begin{equation}
    \begin{aligned}
        &R_2 = z_c \cosh (\tau_c-\Tilde{\tau}_0),\qquad 
         x_0^{(2)} = x_2-R_2,\\ 
        & \tau_c=\Tilde{\tau}_0+\log \bigg( \frac{z_c}{x_2-x_c} \bigg),\qquad 
        \tau_{out} = \Tilde{\tau}_0+ \arccosh \bigg( \frac{z_c^2+|x_2-x_c|^2}{2 \eps |x_2-x_c|} \bigg).
    \end{aligned}
\end{equation}
Equating two expressions for $\tau_c$ and setting $\tau_0=0$, we obtain
\begin{equation}
    \Tilde{\tau}_0 = \log \bigg( \frac{(x_c-x_1)(x_2-x_c)}{z_c^2}  \bigg).
\end{equation}
The bulk actions again vanishes for this configuration, and non-zero contribution to the action comes solely from the boundary terms \eqref{boundary-terms}.
Then, the euclidean action for this configuration is
\begin{multline} \label{2pt_action}
    i S = -k(\tau_c-\tau_{in}) -k (\tau_{out}-\tau_c) = k (\tau_{in} -  \tau_{out}) =-k \bigg[ \log\bigg(  \frac{(x_c-x_1)(x_2-x_c)}{z_c^2}\bigg) +\\+ \arccosh \bigg( \frac{z_c^2+(x_2-x_c)^2}{2 \eps |x_2-x_c|}  \bigg) + \arccosh \bigg( \frac{z_c^2+|x_c-x_1|^2}{2 \eps |x_c-x_1|}\bigg)   \bigg].
\end{multline}
Varying w.r.t. $x_c$ we obtain
\begin{equation}
    \frac{ (x_1+x_2-2 x_c) \left((x_1-x_c) (x_2-x_c)+z_c^2\right)}{\left((x_1-x_c)^2+z_c^2\right) \left((x_2-x_c)^2+z_c^2\right)}=0,
\end{equation}
which is solved by
\begin{equation}
    x_c=\frac{x_1+x_2}{2}.
\end{equation}
Varying w.r.t. $z_c$ we obtain
\begin{equation}
    -\frac{ z_c}{(x_1-x_c)^2+z_c^2}-\frac{ z_c}{(x_2-x_c)^2+z_c^2}+\frac{1}{z_c}=0,
\end{equation}
which is solved by
\begin{equation}
    z_c = \sqrt{(x_c-x_1)(x_2-x_c)} = \frac{x_2-x_1}{2}.
\end{equation}
Then, the worldvolume in AdS becomes just a semicircular geodesic between insertion points $x_1$ and $x_2$, and plugging this solution into the action \eqref{2pt_action} one obtains
\begin{equation}
    e^{iS_{\text{on-shell}}} = e^{-2k \log( {|x_2-x_1|}/{\eps})} = \bigg(\frac{\eps}{|x_1-x_2|}\bigg)^{2k},
\end{equation}
i.e. the variational principle is well-defined and gives the correct coordinate behavior of the two-point function.

Now, what about the giant graviton three-point functions? Let us consider for simplicity the extremal three-point function $\langle \chi_{k_1}(Z) \chi_{k_2}(Z) \chi_{k_1+k_2}(\bar{Z}) \rangle$. The simplification here is that since all three operators depend on the complex scalar $Z$ (and $\bar{Z}$), all three branes created by insertions of these operators wrap the same $S^3 \subset S^5$, and the worldvolume in AdS has a "skeleton" structure of three geodesics meeting at some common point $(\Vec{x}_c,z_c)$, as is shown in Fig. \ref{fig:3gg-worldvolume}:
\begin{figure}[H]
    \centering
    \includegraphics[width=0.6\linewidth]{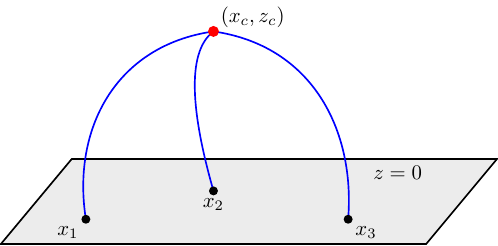}
    \caption{The ansatz for the worldvolume of the three-point function of giant gravitons in $AdS_5$ part. Each line represents trajectory of the D3-brane in AdS, the spatial worldvolumes of the branes are in the $S^5$ part of the spacetime.}
    \label{fig:3gg-worldvolume}
\end{figure}

To simplify the analysis, let us take the geodesic corresponding to the third operator $\chi_{k_1+k_2}(\bar{Z})$ to be a vertical line in Poincar\'e coordinates, i.e. we take the following ansatz
\begin{equation}
    \sigma^0 = t, \qquad \sigma^i = \beta_i, \qquad z=z(t), \qquad x = x_3 = \text{const}. 
\end{equation}

The equation is
\begin{equation}
    \frac{d}{d\tau} \bigg(\frac{\dot{z}}{z^2} \bigg) = - \frac{\dot{z}^2}{z^3},
\end{equation}
which is solved by
\begin{equation}
    z(t) = c_1 e^{c_2 \tau},
\end{equation}
where $c_{1,2}$ are arbitrary constants.
We constrain this solution to satisfy
\begin{equation}
    \frac{\dot{z}^2}{z^2}=1,
\end{equation}
such that the expression for the angular momentum is the same as for the circular geodesic, i.e. $k=N \sin^2\theta$. Then, the solution for the third geodesic is
\begin{equation}
    \begin{aligned}
        z = e^{-(\tau-\tau_0^{(3)})},\qquad 
        \tau_c = \tau_0^{(3)} - \log z_c,\\
        \tau_0^{(3)} = \log (x_3-x_1),\qquad 
        \tau_{out}^{(3)} = \log \frac{x_3-x_1}{\eps}.
    \end{aligned}
\end{equation}
For the first geodesic the solution is as above \eqref{1st-geodesic}, for the second one the solution is
\begin{equation}
    x(\tau)=x_0^{(2)}-R_2 \tanh (\tau-\Tilde{\tau}_0),\qquad 
        z(\tau) = \frac{R_2}{\cosh(\tau-\Tilde{\tau}_0)},
\end{equation}
and from the boundary conditions one obtains
\begin{equation}
    \begin{aligned}
        x_0^{(2)} = x_2-R_2,\qquad 
        R_2 = z_c \cosh (\tau_c-\Tilde{\tau}_0),\qquad 
        \tau_c=\Tilde{\tau}_0+\log \bigg( \frac{x_2-x_c}{z_c} \bigg),\\
        \Tilde{\tau}_0 = \log \bigg( \frac{x_c-x_1}{x_2-x_c} \bigg),\qquad 
        \tau^{(2)}_{in} = \log \bigg( \frac{x_c-x_1}{x_2-x_c}\bigg) - \arccosh \bigg( \frac{(x_2-x_c)^2+z_c^2}{2 \eps (x_2-x_c)}   \bigg),
    \end{aligned}
\end{equation}
where we have already glued this with the solution for the first geodesic and set $\tau_0=0$.

Then, the action is
\begin{multline}
    i S = -k_1 \arccosh \bigg( \frac{(x_3-x_1)^2+z_c^2}{2 \eps (x_3-x_1)}  \bigg) + k_2 \bigg[ \log \bigg( \frac{x_3-x_1}{x_2-x_3} \bigg) - \arccosh \bigg( \frac{(x_2-x_3)^2+z_c^2}{2 \eps (x_2-x_3)}  \bigg)  \bigg] -\\-(k_1+k_2)\log \frac{x_3-x_1}{\eps}. 
\end{multline}
Varying this action w.r.t. $z_c$ one obtains the following equation:
\begin{equation}
    \frac{ k_1 z_c}{(x_1-x_3)^2+z_c^2}+\frac{ k_2 z_c}{(x_2-x_3)^2+z_c^2}=0,
\end{equation}
which is solved by $z_c=0$. Plugging this solution into the action one obtains
\begin{equation}
    e^{i S_{\text{on-shell}}} = e^{-2k_1 \log |{x_{3}-x_1}|/{\eps} -2k_2 \log |{x_{2}-x_3}|/{\eps}} = \bigg(\frac{\eps}{|x_1-x_3|}\bigg)^{2k_1}\bigg(\frac{\eps}{|x_2-x_3|}\bigg)^{2k_2},
\end{equation}
i.e. we see that variational problem for the three-point function configuration gives a saddle which recovers the correct coordinate dependence of the extremal three-point function.

In this subsection we have demonstrated that the boundary terms for the giant gravitons which were discussed in previous subsections, are necessary to reproduce the coordinate dependence for the two-point functions and extremal three-point functions. It will be shown in a future publication, that for the extremal three-point functions of giant gravitons there is a factorization property: dynamics in the $AdS_5$ part (considered here) gives rise to the coordinate dependence, while the dynamics in the $S^5$ part gives large $N$ asymptotics of the structure constant. 


However, let us consider a non-extremal three-point function of three
half-BPS operators, for example\footnote{Here we have assumed that the two-point functions are unit-normalized.}
\begin{equation}
    \langle \chi_{k_1} (y_1 \cdot \Phi(x_1))\, \chi_{k_2} (y_2 \cdot \Phi(x_2)) \,\chi_{k_3} (y_3 \cdot \Phi(x_3)) \rangle = \prod_{i<j}^3 \bigg(\frac{y_i \cdot y_j}{|x_i-x_j|^2} \bigg)^{b_{ij}} C_{k_1k_2k_3},
\end{equation}
where $\Phi=(\Phi_1,\ldots,\Phi_6)$ is the six-vector of scalar
fields of $\mathcal{N}=4$ SYM, the $y_i$ are null six-dimensional
polarization vectors, $C_{k_1k_2k_3}$ is the structure constant\footnote{See \cite{Abajian:2023jye} for explicit expressions for the structure constants for the giant gravitons and dual giant gravitons.}, and $b_{ij}$ are the bridge lengths defined as
\begin{equation}
    b_{ij}=\frac{k_i+k_j-k_\ell}{2},
    \qquad \{i,j,\ell\}=\{1,2,3\}.
\end{equation}
The corresponding polarization vectors may be chosen as
\begin{equation}
    y_1=(1,i,0,0,0,0),\qquad
    y_2=(1,-i,0,0,0,0),\qquad
    y_3=(1,0,i,0,0,0),
\end{equation}
which gives $\langle \chi_{k_1} (Z) \chi_{k_2} (\bar{Z}) \chi_{k_3} ((Z+\bar{Z}+i(Y+\bar{Y}))/2)$.
Note that the third operator here depends on the complex scalar $Z+\bar{Z}+i(Y+\bar{Y})$ as opposed to $Z$ in the extremal case, which means that the third D3-brane wraps another $S^3 \subset S^5$ than the first two D3-branes.
The expected worldvolume structure is again
given in Fig. \ref{fig:3gg-worldvolume}, where the meeting point $(x_c,z_c)$ is found by
varying the action as shown above and is a non-trivial function of
the coordinates $x_{1,2,3}$ and charges $k_{1,2,3}$ (similarly to the case of three-point functions of heavy strings states considered in \cite{Minahan:2012fh,Klose:2011rm}).
The fact that the saddle $(x_c\,,\,z_c)$ in this case also depends on the momenta $k_i$ suggests that, in the non-extremal case, the boundary action may contribute not only to the coordinate dependence but also to the semiclassical structure constant.

A complete computation, however, requires the explicit supersymmetric D3-brane worldvolume, including its dynamics in the internal directions. The details of calculation of the extremal three-point function (including the structure constant) of giant gravitons from the gravity side, as well as non-extremal three-point functions, will be given in a future work.

\section{LLM two-point functions} \label{sec-LLM}

Now, let us derive the two-point functions of two heavy ($\Delta \sim N^2$) operators from 10d supergravity. Notice that in this case one has to evaluate the ten-dimensional pseudo-action of Type IIB supergravity, which is why the approach we take here differs from the one taken in \cite{Skenderis:2007yb} with KK reduction, and is closer in spirit to \cite{Abajian:2023jye}. Below we demonstrate that the same phenomenon happens as for the giant graviton case, i.e. the bulk term in the action vanishes due to supersymmetry, and the on-shell action has contributions only from the boundary terms. In this case the boundary term will be given by the well-known Gibbons-Hawking-York term which makes variation over metric well-defined, which is analogous to the term we have derived in the previous section.

We expect that
\begin{equation}
    \langle \mathcal{O}_{\Delta}(x_1) \mathcal{O}_{\Delta}(x_2) \rangle \sim e^{i S_{\text{on-shell}}^{\text{LLM}}}, 
\end{equation}
where the LLM pseudo-action is
\begin{equation}
    S^{\text{LLM}} = S_{\text{bulk}} + S_{\text{GHY}},
\end{equation}
where
\begin{equation}
    \begin{aligned}
        S_{\text{bulk}} = \frac{1}{2 \kappa^2} \int_{\mathcal{M}} d^{10}x \sqrt{-g} \bigg( R-\frac{1}{4 } F_5^2 \bigg),\\
        S_{\text{GHY}} = -\frac{1}{\kappa^2} \int_{\partial \mathcal{M}} d^9x\bigg(\sqrt{h} \,K -(\sqrt{h} \,K)_{AdS} \bigg).
    \end{aligned}
\end{equation}

For the LLM background the Ricci scalar is identically zero, and the $F_5^2$ term also vanishes due to the self-duality of the five-form. Then, the whole on-shell value will come from the Gibbons-Hawking-York boundary term. Ideally, what we want to obtain is that $S_{\text{on-shell}} = -\Delta \int dt$. To do that, we need asymptotic behavior of the LLM metric near the boundary, i.e. we need to identify the boundary and derive expansion of LLM metric there.

The Lin-Lunin-Maldacena geometry is given by \cite{Lin:2004nb}:
\begin{equation} \label{LLM_metric}
    ds^2 = -h^{-2} (dt + V_i dx^i)^2 + h^2 (dy^2 + dx^i dx^i ) + ye^G d\Omega_3^2 + ye^{-G} d\Tilde{\Omega}_3^2,
\end{equation}
where
\begin{equation}
    h^{-2} = 2y \cosh G, \qquad \zeta = \frac{1}{2} \tanh G, \qquad y \partial_y V_i = \epsilon_{ij} \partial_j \zeta,
\end{equation}
or
\begin{equation} \label{h2m-yeg}
    h^{-2} = \frac{2y}{\sqrt{1-4\zeta^2}}, \qquad e^G = \sqrt{\frac{1+2\zeta}{1-2\zeta}}.
\end{equation}
The function $\zeta(x_1,x_2,y)$ obeys $6d$ Laplace equation, from which one obtains
\begin{align}
    &\zeta(x_1,x_2,y) =\frac{1}{2}- \frac{y^2}{\pi} \int_{\mathcal{D}} \frac{\rho(x') dx'_1 dx'_2}{[(\Vec{x}-\Vec{x}')^2+y^2]^2},\\
    &V_i (x_1,x_2,y)= - \frac{\eps_{ij}}{\pi} \int_{\mathcal{D}} \frac{\rho(x')  (x_j-x'_j)dx'_1 dx'_2}{[(\Vec{x}-\Vec{x}')^2+y^2]^2} ,
\end{align}
where $\rho(x)$ is equal to 1 in the black region (the so-called "droplet", which we denote by $\mathcal{D}$) and 0 in the white. The form of the droplet depends on the operator on the CFT side, and one way to derive the form of the droplet directly from the CFT is to rewrite the two-point function path integral as an integral over a complex matrix $Z$ and find its eigenvalue distribution \cite{Anempodistov:2025maj}. For the finite simply-connected droplets the geometry is asymptotically $AdS$, and the boundary is located at $y^2+x_1^2+x_2^2=R^2 \to \infty$ \cite{Lin:2004nb}.

It will be convenient to write down expansion of the metric in terms of the multipole moments, defined as:
\begin{equation} \label{moments}
    \begin{aligned}
       &M_0 = \frac{1}{\pi} \int_{\mathcal{D}}  d^2x,\\
        &M_i = \frac{1}{\pi} \int_{\mathcal{D}}  x_i \, d^2x,\\
        &M_{ij} = \frac{1}{\pi} \int_{\mathcal{D}}  x_i x_j \, d^2x.
    \end{aligned}
\end{equation}

In the following, we will assume that the droplet is radially symmetric\footnote{For example, one can take the operator to be a Schur polynomial with $\sim N^2$ boxes in the Young tableau.} (which will simplify the calculations), and denote $M \equiv M_{11} = M_{22}$.

First, let us determine the operator dimension in terms of the gravity quantities and multipole moments. Using \cite{Lin:2004nb}
\begin{equation}
    \hbar = 2 \pi l_p^4, \qquad 2\kappa^2 = (2\pi)^7 l_p^8,
\end{equation}
the operator dimension (which is derived from the rotational terms in the LLM metric at the asymptotic infinity \cite{Lin:2004nb}) can be rewritten as
\begin{equation} \label{dimension}
    \Delta = J= \frac{1}{4\hbar^2} \bigg[ \int_{\mathcal{D}} \frac{d^2x}{\pi} (x_1^2+x_2^2)-\frac{1}{2}  \bigg( \int_{\mathcal{D}} \frac{d^2x}{\pi} \bigg)^2  \bigg] = \frac{8\pi^5}{\kappa^2} \bigg( M-\frac{1}{4}M_0^2 \bigg).
\end{equation}

Let us also set the AdS radius to identity, which implies $M_0=1$, and then restore $M_0$ in the end by dimensional analysis. Making a change of variables as follows
\begin{equation} \label{LLM-change-of-variables}
    y = \sqrt{R^2-1} \cos \theta, \qquad x_1 = R \sin \theta \cos \phi , \qquad x_2 = R \sin \theta \sin \phi,
\end{equation}
we take $R \to \infty$ limit and obtain
\begin{multline} \label{z_expansion}
    \zeta(x_1,x_2,y) = \frac{1}{2} - \frac{y^2}{ (r^2+y^2)^2} \int_{\mathcal{D}} \frac{d^2x'}{\pi} \bigg[ 1 - \frac{2}{r^2+y^2} \bigg( \Vec{x}'^2-\frac{6r^2 (\Vec{x}' \Vec{n})^2}{r^2+y^2} \bigg) \bigg] = \\= \frac{1}{2}-\frac{\cos^2 \theta}{R^2} \bigg(  1 - \frac{4M}{R^2} (1-3\sin^2 \theta)\bigg) + O(R^{-6}),
\end{multline}
where we have denoted $r \equiv R \sin \theta$. For $V_i$ we obtain the following multipole expansions:
\begin{equation}
    \begin{aligned}
        &V_1 = - \frac{x_2}{(r^2+y^2)^2} \int_{\mathcal{D}} \frac{d^2x'}{\pi} \bigg[1 - \frac{2 \Vec{x}'^2}{r^2+y^2} + \frac{12 r^2 (\Vec{x}' \Vec{n})^2}{(r^2+y^2)^2}- \frac{4 x_2' x_2'}{r^2+y^2}+...  \bigg],\\
        &V_2 =   \frac{x_1}{(r^2+y^2)^2} \int_{\mathcal{D}} \frac{d^2x'}{\pi} \bigg[1 - \frac{2 \Vec{x}'^2}{r^2+y^2} + \frac{12 r^2 (\Vec{x}' \Vec{n})^2}{(r^2+y^2)^2}- \frac{4 x_1' x_1'}{r^2+y^2}+...  \bigg].
    \end{aligned}
\end{equation}
For the radially symmetric droplet we get
\begin{equation}
    V_r = V_1 \cos{\phi} + V_2 \sin{\phi} =0,
\end{equation}
and
\begin{equation} \label{V-expansion}
    V_{\phi} = -x_2 V_1 + x_1 V_2 = \frac{\sin^2\theta}{R^2} \bigg[ 1 - \frac{4M}{R^2}(2-3\sin^2\theta) +...\bigg]. 
\end{equation}
The relevant part of the metric is
\begin{equation}
    ds^2 = -h^{-2} dt^2 -2h^{-2}V_{\phi}dt d\phi-h^{-2} V_{\phi}^2 d\phi^2+ h^2 (dy^2 + dx^i dx^i) + y e^{G} d\Omega_3^2 + ye^{-G} d \Tilde{\Omega}_3^2,
\end{equation}
which after the change of coordinates $\varphi = \phi-t$ reads
\begin{multline} \label{metric-tilde}
    ds^2 = -(h^{-2}+2h^{-2} V_{\phi} + h^{-2}V_{\phi}^2-h^2 R^2 \sin^2\theta ) dt^2 + 2(h^2 R^2 \sin^2\theta -h^{-2} V_{\phi}^2-h^{-2} V_{\phi}) dt d\varphi + \\+(h^2 R^2 \sin^2 \theta -h^{-2} V_{\phi}^2) d\varphi^2+h^2 R^2 \frac{R^2-\sin^2\theta}{R^2-1} (\frac{dR^2}{R^2} + \frac{R^2-1}{R^2} d\theta^2) + ye^G d\Omega_3^2+ye^{-G} d\Tilde{\Omega}_3^2.
\end{multline}

Writing down the asymptotic expansions for large $R$ (see Appendix \ref{app_asymptotic-expansions}) and plugging them into the metric above, we obtain
\begin{equation}
    \begin{aligned}
         &g_{tt} \to -R^2-\frac{1}{4} (4 M-1) (3 \cos (2 \theta )-1) + O(R^{-2}),\\
        &g_{t \varphi} \to \frac{(4 M-1) \sin ^2(\theta )}{R^2} +O(R^{-4}),\\
        &g_{\varphi\varphi} \to \sin ^2(\theta )-\frac{(4 M-1) \sin ^2(\theta ) (3 \cos (2 \theta )-1)}{4 R^2}+O(R^{-4}),\\
        &g_{RR} \to \frac{1}{R^2}+O(R^{-4}),\\
        &g_{\theta \theta} \to   1+  \frac{\left(\frac{1}{4}-3 M\right) \cos (2 \theta )+M+\frac{1}{4}-\sin ^2(\theta )}{R^2}+O(R^{-4}),
    \end{aligned}
\end{equation}
and expansions for $y e^G$ and $y e^{-G}$ can be found in the Appendix \ref{app_asymptotic-expansions}.

Now we make the change of variables as
\begin{equation}
    z=\frac{1}{R},
\end{equation}
and the boundary $\partial \mathcal{M}$ will be given by a hyperplane $z=\eps$.
The extrinsic curvature is
\begin{equation}
    K =   \frac{1}{2} n^m h^{ab}h_{ab,m} = - \frac{z}{2}h^{ab}h_{ab,z} ,
\end{equation}
where $h_{ab}$ is the induced metric on the boundary.
Plugging here the asymptotic expansions given above we obtain
\begin{equation}
    K = 4+\frac{1}{4}  \bigg( 3 (4 M-1) \cos (2 \theta )-4 M+13 \bigg)\epsilon ^2 + O(\eps^4).
\end{equation}
Calculating the square root of the induced metric, we find that the integrand of the GHY term has the following $\eps$ expansion
\begin{multline} \label{GH_LLM}
     \sqrt{-h}\,K =\bigg[-\frac{3}{32} \bigg(17 \cos (4 \theta )+41 +6 \cos ^2(2 \theta ) -48 M^2 -48 M^2 \cos (4 \theta ) +96 M^2 \cos ^2(2 \theta )-\\-72 M +64 M \cos (2 \theta ) -72 M \cos (4 \theta ) -48 M \cos ^2(2 \theta ) \bigg)+\frac{3 \cos (2 \theta ) -13 +4 M -12 M \cos (2 \theta ) }{4 \epsilon ^2}+\\+\frac{4 }{\epsilon ^4}\bigg] \cos^3\theta \sin \theta  \sin^2\alpha_1 \sin \alpha_2 \sin ^2\beta_1 \sin\beta_2 +O(\eps^2).
\end{multline}

Now we turn to calculation of the GHY term for $AdS_5 \times S^5$ spacetime.
For pure $AdS_5 \times S^5$ case the droplet is given by a disc\footnote{The droplet is a disc with unit radius since we have set $M_0=1$.} \cite{Lin:2004nb} and one has
\begin{equation}
    \begin{aligned}
        &\zeta(r,y) = \frac{r^2-1+y^2}{2\sqrt{(r^2+1+y^2)^2-4r^2 }},\qquad 
        &V_{\phi}(r,y) = \frac{1}{2} \bigg( \frac{r^2+y^2+1}{\sqrt{(r^2+1+y^2)^2-4r^2 }}-1 \bigg). 
    \end{aligned}
\end{equation}
Changing the variables as \eqref{LLM-change-of-variables}
and expanding for large $R$, we obtain
\begin{equation}
    \begin{aligned}
        \zeta = \frac{1}{2} - \frac{\cos^2 \theta}{R^2} - \frac{\sin^2\theta \cos^2 \theta}{R^4} + O(1/R^6),\\
        V_{\phi} = \frac{\sin^2 \theta}{R^2} + \frac{\sin^4 \theta}{R^4}+O(1/R^6).
    \end{aligned}
\end{equation}

We again make the rotation $\varphi=\phi-t$ to put the metric in the form \eqref{metric-tilde} and write down asymptotic expansion for the metric components:
\begin{equation}
    \begin{aligned}
        &g_{tt}\to-R^2+O(R^{-2}),\\
        &g_{t \varphi} \to O(R^{-4}),\\
        &g_{\varphi\varphi} \to \sin^2\theta + O(R^{-4}),\\
        &g_{RR} \to \frac{1}{R^2}+O(R^{-4}),\\
        &g_{\theta \theta } \to 1+O(R^{-4}),\\
        &y e^G \to R^2-1+\frac{\sin ^4(\theta )}{2 R^2}+O(R^{-4}),\\
        &y e^{-G} \to \cos^2 \theta + O(R^{-4}).
    \end{aligned}
\end{equation}
Then, we again make the change of variables $z=1/R$ and calculate the GHY term on the cutoff $z=\eps$. In this case, the integrand of the GHY term is
\begin{equation} \label{GH_AdS}
    (\sqrt{h}K)_{AdS} =\bigg[ \frac{4}{\epsilon ^4}-\frac{3 }{\epsilon ^2}+3 \left(\sin ^4\theta -1\right)  \bigg] \sin^2\alpha_1 \sin\alpha_2 \sin^2\beta_1 \sin\beta_2\sin\theta \cos^3\theta + O(\eps^2).
\end{equation}

Then, if we integrate the difference of \eqref{GH_LLM} and \eqref{GH_AdS} and restore the dimensions, we obtain

\begin{multline}
    S_{\text{GHY}}^{\text{on-shell}} = -\frac{1}{\kappa^2} \int_{\partial \mathcal{M}} d^9x\bigg(\sqrt{h} \,K -(\sqrt{h} \,K)_{AdS} \bigg) = - \frac{8\pi^5}{\kappa^2} \bigg( M-\frac{1}{4}  \bigg) \int dt \rightarrow \\ \rightarrow - \frac{8\pi^5}{\kappa^2} \bigg( M-\frac{M_0^2}{4}  \bigg) \int dt = - \Delta \int dt,
\end{multline}
where we have used expression \eqref{dimension} for the dimension of the operator. 
We therefore find that the on-shell action is entirely determined by the endpoints of the time interval. These endpoints should not be viewed as arbitrary cutoffs: these are the boundary data that encode the positions of the two heavy operator insertions on the boundary, in complete analogy with the discussion of Section~\ref{sec-GG}. Consequently, the regulated on-shell action reproduces the expected spacetime dependence of the heavy two-point function:
\begin{equation}
    \langle \mathcal{O}_{\Delta}(x_1) \mathcal{O}_{\Delta}(x_2) \rangle \sim e^{i S_{\text{on-shell}}^{\text{LLM}}} = \bigg(\frac{\eps}{|x_1-x_2|}\bigg)^{2\Delta}.
\end{equation}
Note that this should be distinguished from the prescription in \cite{Abajian:2023jye,Abajian:2023bqv}, where one attempts to construct a genuinely Euclidean two-point-function geometry whose boundary sources are localized at two separated points. 

Here we have demonstrated that the on-shell value of the Gibbons-Hawking-York term calculated in an arbitrary\footnote{Here for simplicity we have considered radially symmetric droplets, but the analysis can be generalized straightforwardly to any non-symmetric droplet provided it is asymptotically AdS.} LLM background  reproduces the correct coordinate dependence of the two-point function. 
On the gauge theory side one can consider, for example, Schur polynomial operators in huge representations (with number of boxes in Young tableau of order $N^2$), which will be dual to LLM geometries with a droplet given by concentric annuli. However, one can also consider operators that are dual to giant gravitons, which are supposed to be described by the AdS disc droplet with a thin ring excitation \cite{Skenderis:2007yb,Takayama_2005,Anempodistov:2025maj}. In Appendix \ref{app-thin-ring} we demonstrate how giant graviton two-point functions can also be reproduced using the thin ring distributions directly from LLM geometries using the results of this section.

\section{Conclusions}

In this paper we have calculated the two-point function of heavy ($\Delta \sim N$ and $\Delta \sim N^2$) half-BPS operators by evaluating directly the on-shell actions of their holographic duals in the ten-dimensional supergravity. Although the coordinate dependence of the two-point function
can be derived indirectly by determining the energy of the corresponding object in gravity, it is crucial to have direct on-shell action evaluation of this coordinate dependence directly from gravity to have a well-defined setup for gravity calculation of various three-point functions (both the coordinate dependence and structure constants).

For the case $\Delta \sim N$ we have considered the giant graviton operators, and we have shown that in the dual gravity picture the usual D3-brane bulk action should also be endowed with an additional boundary term. The reason for this boundary term to arise is that when one does holographic computation of the two-point function via the path integral \eqref{path-int-2pt}, the field $\varphi(t)$ there is implicitly integrated with Neumann boundary conditions that fix the angular momentum of the brane at its endpoints on the Poincaré boundary, and this boundary term is necessary to compensate the variation of the action and make the saddle-point well-defined. We have derived the form of this boundary term and have shown that its on-shell value reproduces the two-point function of the giant graviton operators. Moreover, we have argued that the same boundary terms arise when one considers $\frac{1}{4}$- and $\frac{1}{8}$-BPS giant graviton configurations for D3-branes. 

The importance of the boundary terms for the giant gravitons which are discussed in Sec. \ref{sec-GG} is not limited only to the two-point function calculation, but we believe that it is a crucial ingredient for reproducing the three-point function of giant gravitons from gravity. Having non-zero on-shell action which reproduces the correct behavior of the two-point functions (with normalization recoverable as in \cite{Holguin:2025dei,Bachas:2024nvh}) gives us, in the spirit of \cite{Janik:2010gc, Minahan:2012fh,Klose:2011rm,Janik:2011bd}, a promising set-up for this problem. In Sec. \ref{sec-GG} we have shown that the on-shell boundary term action reproduces the coordinate dependence of the extremal three-point function, and
for non-extremal three-point functions the same boundary action is expected to contribute also to the structure constant, although a full computation requires the explicit supersymmetric D3-brane worldvolume and is left for future work.

For the case $\Delta \sim N^2$ we have considered Schur polynomial operators (with number of boxes in the Young tableau of order $N^2$) on the CFT side, which correspond to Lin-Lunin-Maldacena bubbling geometries with circularly symmetric droplets. In this case we have evaluated the Type IIB supergravity on-shell pseudo-action and have encountered a similar phenomenon that the bulk term in the action vanishes and the only contribution comes from the boundary Gibbons-Hawking-York term. We have written down the asymptotic expansion of the LLM metric near the boundary, evaluated regularized GHY term, and found that it gives the correct behavior of the two-point function. This calculation provides us an intuition on where the coordinate dependence of the correlators comes from when one evaluates the on-shell action. 
An interesting question is whether it is possible to reproduce the normalization of the two-point functions from the supergravity calculation \cite{Gentle:2015ruo, IzquierdoGarcia:2025jyb, Kurlyand:2022vzv, Henneaux:1988gg}.
It would be interesting to understand whether additional finite or topological terms in the renormalized ten-dimensional action can reproduce this normalization, but we do not address this problem here.

It is also important to distinguish the ten-dimensional computation performed here from the lower-dimensional effective action obtained after Kaluza-Klein reduction. In the ten-dimensional type IIB pseudo-action the bulk contribution vanishes on-shell for the half-BPS LLM backgrounds: the Ricci scalar is zero and the self-duality of the five-form implies $F_5^2=0$. Hence the only contribution to the universal classical part of the answer comes from the regulated gravitational boundary term. This is the ten-dimensional analogue of the phenomenon encountered for giant gravitons, where the DBI and Wess-Zumino terms cancel on-shell and the two-point function is produced by the boundary contribution required by the variational problem.

This statement does not imply that the full holographic renormalization problem is trivial. In particular, if one first reduces the solution to a lower-dimensional effective theory, the reduced action may contain divergent terms even though the original ten-dimensional pseudo-action vanishes on-shell. This apparent mismatch is related to the standard subtleties of type IIB supergravity with a self-dual five-form: the pseudo-action is not an ordinary covariant action for unconstrained fields, and dimensional reduction, imposing self-duality, adding boundary terms, and performing holographic renormalization do not necessarily commute. 
For example, the on-shell value of the five-dimensional truncated action evaluated on $AdS_5$ is nonzero and reproduces the planar contribution to the UV-divergent free energy of $\mathcal{N}=4$ SYM on $S^4$. 
Reconciling this result with the vanishing of the ten-dimensional type IIB pseudo-action requires the inclusion of additional boundary terms, as discussed in \cite{Kurlyand:2022vzv}. The purpose of the present computation is more limited. We extract the universal logarithmic dependence on the cutoff, equivalently the universal coordinate dependence of the heavy two-point function. The motivation for this calculation is to set the stage for various three-point functions from the 10d supergravity. For example one can consider in the gauge theory a three-point function of Schur polynomials, where two of them have $\Delta \sim N^2$, and the third is in a single row/column representation with $k \sim N$ boxes. This three-point function, in principle, should be reproduced from the gravity by considering giant graviton D3-branes embedded in the LLM background. Although finding actual supersymmetric embeddings of giant gravitons is a difficult task, the results of this paper give an insight on where do non-zero contributions to the on-shell actions come from and how the variational problem is posed.

Also, let us stress that the remaining finite, scheme-dependent, and operator-dependent normalization of the LLM two-point function is not determined by the Gibbons-Hawking-York term. Of course, this normalization depends on the operator details, or equivalently on the detailed shape of the LLM droplet. Determining this full normalization directly from the renormalized ten-dimensional or reduced action is a separate problem, which we leave for future work.

In fact, using this method we can also calculate two-point functions of operators with $\Delta \sim N$, for example giant gravitons, as is shown in the Appendix \ref{app-thin-ring}. This opens a new perspective on calculation of three-point function where two operators have $\Delta \sim N^2$ and one has $\Delta \sim N$: it is interesting to see whether this kind of three-point function can be reproduced by evaluating the on-shell action for the LLM geometry with the droplet corresponding to two $\Delta \sim N^2$ operators with a thin line excitation created by the third operator. We leave this calculation, as well as calculation of three-point functions of giant gravitons for future work.

\section*{Acknowledgements}
I would like to thank Romuald Janik, Harish Murali, Bogdan Stefanski, and Pedro Vieira for useful discussions. I am especially grateful to Adolfo Holguin and Vladimir Kazakov for discussions that inspired this work and for their insightful comments and suggestions on the draft, and to Constantin Bachas for careful reading of the manuscript and valuable feedback.

\begin{appendices}
    \setcounter{equation}{0}

    \section{Review of the giant graviton two-point function calculation in the previous literature} \label{app-GG}

    Let us first repeat and elaborate on the arguments of \cite{Bissi:2011dc} and then show where they break down.

Introducing an einbein, the action \eqref{GG-action} can be equivalently rewritten as
\begin{equation}
   S_{D3} = \frac{1}{2} \int dt \bigg( \frac{1}{e} g_{\mu \nu} \dot{x}^{\mu} \dot{x}^{\nu} -m^2 e + \frac{1}{e} \cos^2 \theta \, \dot{\varphi}^2 +2N \sin^4 \theta \, \dot{\varphi} \bigg),
\end{equation}
where $g_{\mu \nu}$ is the metric on $AdS_5$ part, $m\equiv N \sin^3 \theta$ and the on-shell value of $e$ is 
\begin{equation}
    e = \frac{1}{m} \sqrt{-g_{\mu \nu} \dot{x}^{\mu} \dot{x}^{\nu} - \cos^2 \theta \, \dot{\varphi}^2}.
\end{equation}
The conjugate momentum to $\varphi$ is
\begin{equation}
    k = \frac{1}{e} \dot{\varphi} \cos^2 \theta + N \sin^4 \theta .
\end{equation}
is minimized for the following value of $\theta$:
\begin{equation}
    \sin^2 \theta_0 = \frac{k}{N}.
\end{equation}
Then, substituting here $k$ we find
\begin{equation}
    \dot{\varphi} = e N \sin^2 \theta_0.
\end{equation}
Plugging this value of the $\dot{\varphi}$ into the action, one finds that last three terms in the action become effective mass term:
\begin{equation}
    -m^2 e + \frac{1}{e} \cos^2 \theta \dot{\varphi}^2+ 2N \sin^4 \theta \dot{\varphi}=e N^2 \sin^4 \theta_0 = e k^2,
\end{equation}
and the action is
\begin{equation}
    S = \frac{1}{2} \int dt \bigg( \frac{1}{e} g_{\mu \nu} \dot{x}^{\mu} \dot{x}^{\nu} +  k^2 e \bigg),
\end{equation}
i.e. we obtain a Lagrangian of a point particle moving in $AdS_5$ with mass equal to the angular momentum $k$ of the giant graviton. 
Now, let's do a Wick rotation $\tau = -i t$ and plug $\rho=0$ ansatz into the action. We obtain
\begin{equation}
    S = \frac{i}{2} \int d\tau \bigg( \frac{1}{e}+  k^2 e \bigg).
\end{equation}
Solving for the einbein, we obtain on-shell
\begin{equation}
    S = \frac{i}{2} \int d\tau \, 2 k.
\end{equation}
Then, repeating the arguments from Sec.\ref{sec-GG} one obtains correct behavior of the two-point function:
\begin{equation}
    e^{iS} \to e^{-S_E} = e^{-2 k \log(|x_1-x_2|/\eps)} = \bigg( \frac{\eps}{|x_1-x_2|} \bigg )^{2k}.
\end{equation}

The error in the arguments above is that the system of equations for $\varphi(t)$ and $e(t)$ are coupled, and one should solve these equations simultaneously. In the arguments above, one solves these equations, and then leaves $\varphi$ as a function of $e$, then plugs this back into the action and writes down saddle-point equation for $e$ again.

Solving the equations
\begin{equation}
    \begin{aligned}
        k = \frac{1}{e} \cos^2 \theta \dot{\varphi} + N \sin^4 \theta, \\
        e = \frac{1}{m} \sqrt{1- \cos^2 \theta \dot{\varphi}^2},
    \end{aligned}
\end{equation}
one obtains the following solution
\begin{equation}
    \dot{\varphi}=1, \qquad e = \frac{1}{N \sin^2 \theta},
\end{equation}
and plugging this back into the action one obtains
\begin{equation}
    S_{\text{on-shell}} = \frac{1}{2} \int dt (-N \sin^2 \theta \sin^2 \theta -N \sin^4 \theta) + N \sin^4 \theta \int dt = 0,
\end{equation}
which agrees with the calculation without einbein:
\begin{multline}
    S_{D3}^{\text{on-shell}} = -N \sin^3 \theta \int dt \sqrt{1- \cos^2 \theta \dot{\varphi}^2}+N\sin^4 \theta \int \dot{\varphi} dt = \\=-N \sin^4 \theta \int dt + N \sin^4 \theta \int dt = 0.
\end{multline}

\section{Giant gravitons in Poincar\'e coordinates} \label{app-gg-eom-poincare}

It is instructive to redo the analysis of the giant graviton solutions in the Poincar\'e coordinates. Making the change of variables
\begin{equation}
    \begin{aligned}
        &z = \frac{1}{\cosh \rho \cos t - n_0 \sinh \rho},\\
        &x^0 = \frac{\cosh \rho \sin t}{\cosh \rho \cos t - n_0 \sinh \rho},\\
        &\Vec{x} = \frac{\Vec{n} \sinh \rho}{\cosh \rho \cos t - n_0 \sinh \rho},
    \end{aligned}
\end{equation}
where $(n_0,\Vec{n}) = (\cos \alpha_1 \,, \, \sin \alpha_1 \cos \alpha_2 \,, \, \sin \alpha_1 \sin \alpha_2 \cos \alpha_3 \,, \, \sin \alpha_1 \sin \alpha_2 \sin \alpha_3)$,
we arrive at the following metric
\begin{equation}
\begin{aligned}
       ds^2& = \frac{1}{z^2}( dz^2 - (dx^0)^2 + (dx^i)^2 )+\\
       +& d\theta^2 +\cos^2 \theta d \varphi^2 + \sin^2\theta\left(d\beta_1^2 + \sin^2 \beta_1 \left(d\beta_2^2 + \sin^2\beta_2 d\beta_3^2\right)\right).
\end{aligned}
\end{equation}
Now, let us choose the following embedding for the brane
\begin{equation}
    \begin{aligned}
        \sigma^0 \equiv \sigma, \qquad \sigma^i = \beta_i, \\
        z=z(\sigma), \qquad x^0 = x^0 (\sigma), \qquad x^i = x^i (\sigma),\\
        \theta=\theta_0, \qquad \varphi = \varphi(\sigma).
    \end{aligned}
\end{equation}
Then, the action reads
\begin{multline}
    S_{D3}= - \frac{N}{2\pi^2} \int d^4 \sigma \bigg( \sqrt{\frac{1}{z^2}(\dot{x}_0^2 - \dot{z}^2-\dot{\Vec{x}}^2)-\cos^2 \theta \dot{\varphi}^2}  \sin^3 \theta - \dot{\varphi} \sin^4 \theta \bigg) \sin^2 \beta_1 \sin \beta_2 = \\ =- N \int d \sigma \bigg( \sqrt{\frac{1}{z^2}(\dot{x}_0^2 - \dot{z}^2-\dot{\Vec{x}}^2)-\cos^2 \theta \dot{\varphi}^2}  \sin^3 \theta - \dot{\varphi} \sin^4 \theta \bigg).
\end{multline}
Note that in this action there is a reparametrization symmetry under $\sigma \to f(\sigma)$. We fix this gauge freedom by identifying $\sigma$ with the global time $t$, so that we have $\dot{\varphi}=1$ on-shell. This will also be more convenient to interpret the structure of the worldsheet in the global coordinates. 

The conserved angular momentum is
\begin{equation}
    k = \frac{\partial L}{\partial \dot{\varphi}} = N \bigg( \frac{\cos^2 \theta \sin^3 \theta \dot{\varphi}}{\sqrt{(\dot{x}_0^2-\dot{z}^2-\dot{\Vec{x}}^2)/z^2- \cos^2 \theta \dot{\varphi}^2}}+ \sin^4 \theta  \bigg),
\end{equation}
and equations of motion are as follows:
\begin{equation}
    \begin{aligned}
        \frac{d}{dt} \bigg( \frac{\dot{x}_0}{z^2 \sqrt{(\dot{x}_0^2-\dot{z}^2-\dot{\Vec{x}}^2)/z^2- \cos^2 \theta \dot{\varphi}^2}}   \bigg)=0,\\
         \frac{d}{dt} \bigg( \frac{\dot{\Vec{x}}}{z^2 \sqrt{(\dot{x}_0^2-\dot{z}^2-\dot{\Vec{x}}^2)/z^2- \cos^2 \theta \dot{\varphi}^2}}   \bigg)=0,\\
          \frac{d}{dt} \bigg( \frac{\dot{z}}{z^2 \sqrt{(\dot{x}_0^2-\dot{z}^2-\dot{\Vec{x}}^2)/z^2- \cos^2 \theta \dot{\varphi}^2}}   \bigg)= \frac{\dot{x}_0^2-\dot{z}^2-\dot{\Vec{x}}^2}{z^3 \sqrt{(\dot{x}_0^2-\dot{z}^2-\dot{\Vec{x}}^2)/z^2- \cos^2 \theta \dot{\varphi}^2}} .
    \end{aligned}
\end{equation}
They can be simplified significantly if we make a reasonable guess that $\dot{\varphi}=1$ along the whole worldvolume. Then, from the expression for the angular momentum we see that $\sqrt{(\dot{x}_0^2-\dot{z}^2-\dot{\Vec{x}}^2)/z^2- \cos^2 \theta \dot{\varphi}^2}$ is constant as a function of time $t$, and the equations of motion are
\begin{equation}
    \begin{aligned}
        \frac{d}{dt} \bigg( \frac{\dot{x}_0}{z^2}   \bigg)=0,\\
        \frac{d}{dt} \bigg( \frac{\dot{\Vec{x}}}{z^2}   \bigg)=0,\\
        \frac{d}{dt} \bigg( \frac{\dot{z}}{z^2}   \bigg)= \frac{\dot{x}_0^2-\dot{z}^2-\dot{\Vec{x}}^2}{z^3},
    \end{aligned}
\end{equation}
i.e. we obtain geodesic equations in AdS. 

Since the geodesic can be obtained as a section of the AdS hyperboloid by a plane, the geodesic will obey the equation
\begin{equation}
    z^2 + \eta_{\mu \nu} (x^{\mu}-X^{\mu})(x^{\nu}-X^{\nu})=R^2,
\end{equation}
where $X^{\mu}$ and $R$ are arbitrary constants.
After Wick rotation
\begin{equation}
    \tau = -it, \qquad x=-ix_0,
\end{equation}
the geodesic equation becomes
\begin{equation}
    z^2+(\Vec{x}-\Vec{X})^2=R^2.
\end{equation}

    \section{Asymptotic expansions} \label{app_asymptotic-expansions}

    Here we write down large $R$ asymptotic expansions for the various functions present in the LLM metric \eqref{LLM_metric}.
    Plugging the multipole expansions \eqref{z_expansion} and \eqref{V-expansion} into the formulae \eqref{h2m-yeg}, we obtain:
\begin{equation}
    h^2 \to \frac{1}{R^2}+\frac{\left(\frac{1}{4}-3 M\right) \cos (2 \theta )+M+\frac{1}{4}}{R^4}+O(R^{-6}),
\end{equation}
and
\begin{multline}
    h^{-2} \to R^2-M-\frac{1}{4}+\left(3 M-\frac{1}{4}\right) \cos (2 \theta )+ \frac{1}{64 R^2} \bigg(-4 (4 M+1) (36 M-7) \cos (2 \theta )+\\+(24 (18 M-7) M+19) \cos (4 \theta )+24 (22 M-5) M+9 \bigg)+O(R^{-4}).
\end{multline}
\begin{multline}
    y e^G \to R^2-M-\frac{3}{4}+\left(3 M-\frac{3}{4}\right) \cos (2 \theta )+\\+\frac{(12 M-1) ((36 M-7) \cos (4 \theta )+44 M-5)+4 (8 M (1-18 M)+3) \cos (2 \theta )}{64 R^2}+O(R^{-4}),
\end{multline}
\begin{multline}
    ye^{-G} \to \cos ^2(\theta )+\\+\frac{\cos (\theta ) \left(\cos ^3(\theta )+\cos (2 \theta ) \cos (\theta )-\cos (\theta )+2 M \cos (\theta )-6 M \cos (2 \theta ) \cos (\theta )\right)}{2 R^2} +O(R^{-4}).
\end{multline}

For the pure AdS case one has
\begin{equation}
    \begin{aligned}
        h^{-2} \to R^2-\sin ^2(\theta )+\frac{\sin ^4(\theta )}{2 R^2} + O(R^{-4}),\\
        h^2 \to \frac{1}{R^2}+\frac{\sin ^2(\theta )}{R^4} +O(R^{-6}).
    \end{aligned}
\end{equation}

\section{Thin ring calculation} \label{app-thin-ring}

\begin{figure}[H]
\centering
\begin{minipage}{.45\textwidth}
  \centering
  \includegraphics[width=.6\linewidth]{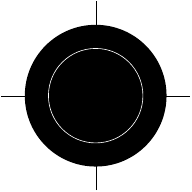}
  \captionof{figure}{LLM droplet distribution corresponding to the $S^5$ giant graviton.}
  \label{fig:gg-droplet}
\end{minipage}%
\hspace{1cm}
\begin{minipage}{.45\textwidth}
  \centering
  \includegraphics[width=.6\linewidth]{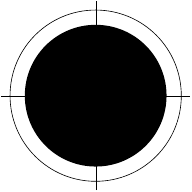}
  \captionof{figure}{LLM droplet for dual giant graviton, which wraps $S^3 \subset AdS_5$.}
  \label{fig:dual-gg-droplet}
\end{minipage}
\end{figure}

It was suggested in the original LLM paper \cite{Lin:2004nb} that the giant gravitons and dual giant gravitons should correspond to LLM distributions consisting of discs with small droplet excitations. However, calculations in \cite{Takayama_2005, Skenderis:2007yb}, as well as a matrix model calculation in \cite{Anempodistov:2025maj}, suggest that they rather should correspond to radially symmetric distributions with thin ring excitations as in Figures \ref{fig:gg-droplet} and \ref{fig:dual-gg-droplet}. Here, using the results of Section \ref{sec-LLM}, we demonstrate that the thin ring distributions reproduce correctly the (dual) giant graviton two-point functions, which supports the latter viewpoint.

We will derive that the conformal dimension, defined through the moments of distribution as
\begin{equation}
    \Delta = \frac{8\pi^5}{\kappa^2} \bigg( M-\frac{1}{4}M_0^2 \bigg),
\end{equation} gives the correct conformal dimension of (dual) giant gravitons for the thin ring distributions.

Let us consider area-preserving deformations of the vacuum droplet density:
\begin{equation}
    \rho = \rho_{disc} + \delta \rho,
\end{equation}
where
\begin{equation}
    \delta M_0 = \frac{1}{\pi} \int \delta \rho \, d^2x =0.
\end{equation}
Since the vacuum (disc droplet) has zero dimension, the conformal dimension of this configuration will be given by
\begin{equation}
    \Delta = \frac{8\pi^5}{\kappa^2} \delta M.
\end{equation}
Let us set $M_0=1$, which amounts to setting AdS radius to one $R_{AdS}=1$. Now, since $R_{AdS}^4 = 4\pi N l_p^4$, one gets
\begin{equation}
    l_p^4 = \frac{1}{4\pi N}, \qquad \hbar = 2\pi l_p^4 = \frac{1}{2N}, \qquad 2\kappa^2 = (2\pi)^7 l_p^8 = \frac{8\pi^5}{N^2}.
\end{equation}
Then, the conformal dimension can be rewritten as
\begin{equation}
    \Delta = 2N^2 \delta M.
\end{equation}
Now let us consider small patch of droplet of area $q$. For the thin ring of area $q$, its contribution to $M$ is
\begin{equation}
    M = \frac{q \,r^2}{2}.
\end{equation}
Taking out a thin ring at a radius $r_-$ and placing it at a radius $r_+$ gives
\begin{equation}
    \delta M = \frac{q}{2} (r_+^2-r_-^2). 
\end{equation}
From  \cite{Anempodistov:2025maj}, the saddle point gives
\begin{equation}
    r_i^2 = \frac{N-i+\lambda_i}{N},
\end{equation}
it is clear that giant graviton thin ring has $q=1/N$, and the conformal dimension can be written as
\begin{equation}
    \Delta = N (r_+^2-r_-^2).
\end{equation}

The giant graviton amounts to taking a thin ring at $r_-^2 = 1 - \frac{k}{N} = \cos^2 \theta$ and placing it at $r_+^2=1$. Plugging this into $\Delta$ one obtains
\begin{equation}
    \Delta = N \bigg(1 - \bigg(1-\frac{k}{N}  \bigg)  \bigg)=k = N \sin^2\theta.
\end{equation}

 For the dual giant graviton one has $r_-^2=1$ and $r_+^2 = 1 + \frac{k}{N} = \cosh^2\rho$, and the conformal dimension is
 \begin{equation}
     \Delta = N \bigg(1 + \frac{k}{N}-1  \bigg)=k = N \sinh^2\rho.
 \end{equation}

Then, evaluation of the GHY term on the background created by both of these distributions reproduces the correct coordinate behavior of the giant graviton and dual giant graviton two-point function.

\end{appendices}

\bibliographystyle{JHEP}
\bibliography{Heavy}

\end{document}